\documentclass[conference]{IEEEtran}
\IEEEoverridecommandlockouts
\ifCLASSOPTIONcompsoc
  
  \usepackage[nocompress]{cite}
\else
  \usepackage{cite}
\fi

\usepackage{cite}
\usepackage{amsmath}
\usepackage{amssymb}
\usepackage{amsfonts}
\usepackage{tikz}

\usepackage{graphicx}
\usepackage{textcomp}
\usepackage{xcolor}
\usepackage{multirow}
\usepackage{soul}
\usepackage{algorithm}
\usepackage{algpseudocode}
\usepackage{colortbl}
\usepackage{bm}
\usepackage{float}
\usepackage{filecontents}
\usepackage{CJK}
\usepackage{algorithm}
\usepackage{algorithmicx}

\usetikzlibrary{automata, arrows.meta, positioning}

\newcommand\encircle[1]{%
\tikz[baseline=(X.base)] 
  \node (X) [draw, scale=0.75, shape=circle, inner sep=0, fill=black, text=white, minimum size=0em] {\strut #1};}

\newcommand{\cmmnt}[1]{}  

\def\BibTeX{{\rm B\kern-.05em{\sc i\kern-.025em b}\kern-.08em
    T\kern-.1667em\lower.7ex\hbox{E}\kern-.125emX}}

\ifCLASSINFOpdf
\else
\fi

\usepackage{todonotes}

\usepackage[a4paper, total={184mm,239mm}]{geometry}
\def\BibTeX{{\rm B\kern-.05em{\sc i\kern-.025em b}\kern-.08em
    T\kern-.1667em\lower.7ex\hbox{E}\kern-.125emX}}

\definecolor{mygreen}{RGB}{9, 181, 55}

\begin{document}
\newcommand{\algrule}[1][.2pt]{\par\vskip.5\baselineskip\hrule height #1\par\vskip.5\baselineskip}

\title{\huge DIAC: \underline{D}esign Exploration of \underline{I}ntermittent-\underline{A}ware \underline{C}omputing Realizing Batteryless Systems\vspace{-0.5em}}

\author{Sepehr Tabrizchi$^\ast$, Shaahin Angizi$^\dagger$, Arman Roohi$^\ast$\\
\small
$^\ast$School of Computing, University of Nebraska–Lincoln, Lincoln NE, USA\\
$^\dagger$Department of Electrical and Computer Engineering, New Jersey Institute of Technology, Newark, NJ, USA\\
aroohi@unl.edu
\vspace{-1.75em}
}

\maketitle
\begin{abstract}
Battery-powered IoT devices face challenges like cost, maintenance, and environmental sustainability, prompting the emergence of batteryless energy-harvesting systems that harness ambient sources.
However, their intermittent behavior can disrupt program execution and cause data loss, leading to unpredictable outcomes.
Despite exhaustive studies employing conventional checkpoint methods and intricate programming paradigms to address these pitfalls, this paper proposes an innovative systematic methodology, namely \texttt{\textbf{DIAC}}. 
The DIAC synthesis procedure enhances the performance and efficiency of intermittent computing systems, with a focus on maximizing forward progress and minimizing the energy overhead imposed by distinct memory arrays for backup. 
Then, a finite-state machine is delineated, encapsulating the core operations of an IoT node, sense, compute, transmit, and sleep states. First, we validate the robustness and functionalities of a DIAC-based design in the presence of power disruptions.
DIAC is then applied to a wide range of benchmarks, including ISCAS-89, MCNS, and ITC-99. The simulation results substantiate the power-delay-product (PDP) benefits. For example, results for complex MCNC benchmarks indicate a PDP improvement of 61\%, 56\%, and 38\% on average compared to three alternative techniques, evaluated at 45 nm.
\vspace{-0.9em}
\end{abstract}

 \vspace{1em}

\section{Introduction}
Shifting from a cloud-centric approach to a thing-/data-centric perspective, known as the Internet of Things (IoT), could significantly mitigate challenges such as high latency, limited scalability, quality of service, privacy, and security.  Given its promising potential, the IoT market is expected to reach \$4.5 trillion by 2035, with an interconnected network of over one trillion devices encompassing smart homes, cities, industries, healthcare wearables/implants, and agriculture \cite{IOT1,hsu2019ai,liu2020energy}. 
According to Ericsson, intelligent IoT systems could reduce carbon emissions by 3\%, or 63.5 gigatons, by 2030 \cite{malmodin2015exploring}.
However, IoT devices are primarily powered by batteries, which presents critical challenges related to limited lifespan, cost, maintenance, and environmental sustainability. EnABLES forecasts that without change, global battery disposal could reach 78 million units daily by 2025 \cite{hayes2019enables}. 

In contrast, batteryless devices can overcome these significant and inherent limitations using energy-harvested systems, which harness ambient energy sources, eliminating the need for traditional batteries. 
However, intermittent behaviors caused by the sources can disrupt program execution and lead to data loss and unpredictable outcomes.
Therefore, advanced techniques in the normally-off (intermittent) computing domain have been formulated. These methodologies offer advantageous features such as near-zero power consumption during idle states, immediate wake-up capabilities, and robustness against power failures \cite{NVM1,Roohi2018TC}. Consequently, non-volatile (NV) components, including non-volatile memories (NVMs) and non-volatile flip-flops (NV-FFs), have received significant industry and academic attention \cite{NVFF1}. These NV elements prevent the need for a boot-up sequence post-sleep owing to their inherent nonvolatility. Typically, in NV processors, data from all registers are offloaded to NVMs before entering a deep sleep state. Throughout this sleep phase, there is no need for a continuous power source.
Upon re-energizing the processor, data are retrieved from the NVMs, and system operations recommence. Recent literature has highlighted various hardware-assisted strategies tailored explicitly for intermittent computing paradigms. The authors of \cite{NVP3} have replaced all traditional flip-flops with NV-FFs, whereas in \cite{FRAM1}, multiple diminutive NVM arrays are used for efficient backup and data restoration. While NVMs present the benefit of data persistence, this comes at the expense of elevated write power consumption. As a result, there is a pressing need for a holistic and systematic synthesis approach.
Traditional checkpoint-based strategies are susceptible to internal and external inconsistencies in the event of a power failure. Internal inconsistencies manifest when the execution context is only partially retained in the NVM, while external inconsistencies occur when power failures happen between successive checkpoints. 
Previous implementations have suffered notably from the computational and energy burdens of additional middleware and checkpointing operations \cite{mementos,DINO}. Furthermore, these traditional methods are prone to leakage between checkpoint operations, particularly when utilizing volatile registers and flip-flops in CMOS-only datapath designs. These challenges necessitate a more comprehensive focus on life cycle energy optimization, considering intermittent power supply and data streams.
Despite exhaustive studies employing conventional checkpoint methods and intricate programming paradigms to address intermittent computing pitfalls, our previous work reveals that such techniques frequently encounter performance bottlenecks and limitations in scalability \cite{Roohi2018TC}. 
\vspace{-0.3em}

This paper developed a cross-layer approach to address the stated issues, which requires effective task scheduling and power management systems.
At the circuit level, power management is designed, establishing various thresholds to delineate different operational zones. At the architecture level, a systematic \underline{D}esign Exploration of \underline{I}ntermittent-\underline{A}ware \underline{C}omputing (\texttt{\textbf{DIAC}}) methodology is devised, which minimizes overhead while maximizing forward computational capabilities. The prototyped DIAC design tool, when integrated with the proposed finite-state machine, offers the core operations of an IoT node, sense, compute, transmit, and sleep states, ensuring resilience against power interruptions. By collaborating software and hardware stages, the number of checkpoints is reduced, reducing both power and delay overhead. In the final phase, the system is deployed in a power-scarce environment, and its performance is evaluated using a system-level in-house framework, including a set of benchmarks.

\section{Intermittent Computing Review}
Ambient energy sources like solar, kinetic, etc., play a vital role in pursuing sustainable and renewable energy solutions. They offer benefits such as abundance, decreased carbon emissions, and cost savings\cite{akella2009social}. 
When considering embedded systems, radio-frequency identification (RFID) is often preferred as an ambient power source for several reasons, including its self-powering mechanism, scalability, and cost-efficiency.
RFID technology uniquely draws power wirelessly from its reader, a feature facilitated by electromagnetic induction\cite{ferdous2016renewable}. Unlike conventional devices that require internal batteries or wired connections, when an RFID reader emits a signal, it induces a current in the RFID tag's antenna. This powers the tag, allowing data transmission without needing batteries or regular maintenance. Although this offers efficiency in applications such as inventory management and access control, intermittent energy bursts can cause operational interruptions, leading to data loss or inconsistent results\cite{eriksson2007mspsim}.
Our research focused on designing a specialized architecture using RFID sources, making integration into embedded systems more straightforward. Based on the observed behavior of this energy source, we take into account its intermittency and variability. Our objective is to optimize energy utilization and minimize dependency on nonrenewable resources by carefully designing our system to adapt to voltage-level fluctuations.

Energy harvesting devices employ intermittent computing, where short bursts of program execution are often interrupted by power failures. Despite its challenges, many studies have explored techniques to leverage this computing style at various levels. Certain challenges must be considered during the design time; the less energy used for computation, the more tasks can be performed. Task execution times can vary, with compute-intensive tasks sometimes interrupted by power losses, potentially rendering data or computations obsolete before completion. System designers need robust language and software support to interact with sensors despite fluctuating power \cite{hester2017future}. They also need a deep understanding of the hardware and behavior of energy harvesters.
From a software-level point of view, intermittent computing involves using checkpoints and task-based programming models \cite{umesh2021survey,singla2022survey}.
Checkpointing enables the preservation of volatile states in non-volatile memory (NVM), allowing computation to resume after a power failure. Meanwhile, task-based systems break programs into short tasks and resume execution of the last completed task. Task-based systems often employ various memory management techniques, such as privatization and variable copying.
 However, checkpoint approaches may suffer from internal and external inconsistencies after each power loss, and frequent checkpointing results in performance degradation.
Furthermore, there are hardware modifications and approaches based on speculation and watchdog timers to optimize data storage and restore tasks during power failures. 
Programming languages like Chain \cite{colin2016chain} and Mayfly \cite{hester2017timely} simplify intermittent computing by segmenting programs into tasks and managing time and memory consistency.
Compiler-based approaches analyze program sections and determine the minimum number of checkpoints needed to prevent incompatibilities. 
The main disadvantages of this model are its incompatibility with existing libraries and code bases, its lack of importability across hardware platforms, and the burden it places on the programmer.
Architecture-level techniques for intermittent computing offer various solutions to overcome the challenges of power failures and optimize energy efficiency \cite{NORM}. Approximate computing techniques are used to reduce the accuracy of computations to improve performance and efficiency. 
Hardware solutions such as Clank track memory access to detect write-after-read sequences, while Cascaded Hierarchical Remanence Timekeeper (CHRT) provides resilient timekeeping during power failures \cite{de2020reliable}. 


\section{Proposed Intermittent-Aware System}
\subsection{Hierarchical Design Exploration}
\begin{figure}[t]
  \centering
    \includegraphics[width=0.7\linewidth]{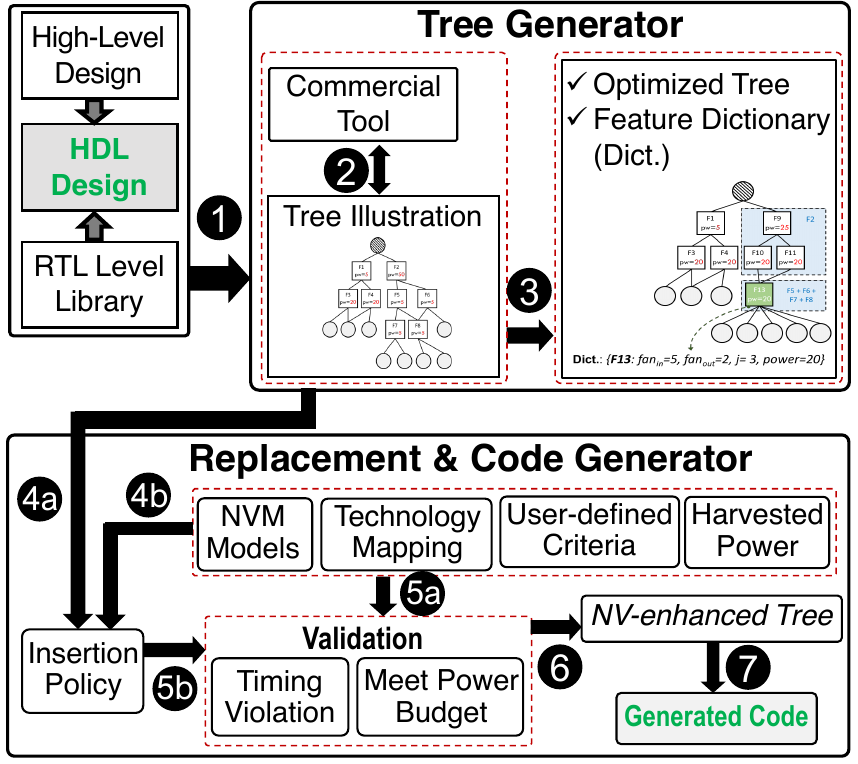}
\vspace{-0.75em}
    \caption{\small Design flow of the DIAC framework.}
    \vspace{-1.75em}
\label{auto}
\end{figure}

Since the charge and discharge cycle – an intrinsic characteristic of energy harvesting devices – may occur more than hundreds of times per second, NV elements can maintain their computation state and ensure forward progress in the presence of unpredictable energy failures.
Critical difficulties encountered in prior research include the requirement for intricate software that lacks scalability, performance overhead introduced by NVM components, and system inconsistencies.
In this section, a co-design methodology is introduced to push the boundaries of intermittent computing. This methodology entails the development of a novel architecture and software strategy that minimizes overhead while maximizing forward progress.
Figure \ref{auto} shows the design flow of our systematic \underline{D}esign Exploration of \underline{I}ntermittent-\underline{A}ware \underline{C}omputing (\texttt{\textbf{DIAC}}) methodology to address power failure with minimum performance overhead, offered by three complementary procedures:


\subsubsection{Tree generator}
\begin{figure}[t]
  \centering
    \includegraphics[width=0.7\linewidth, height=7cm]{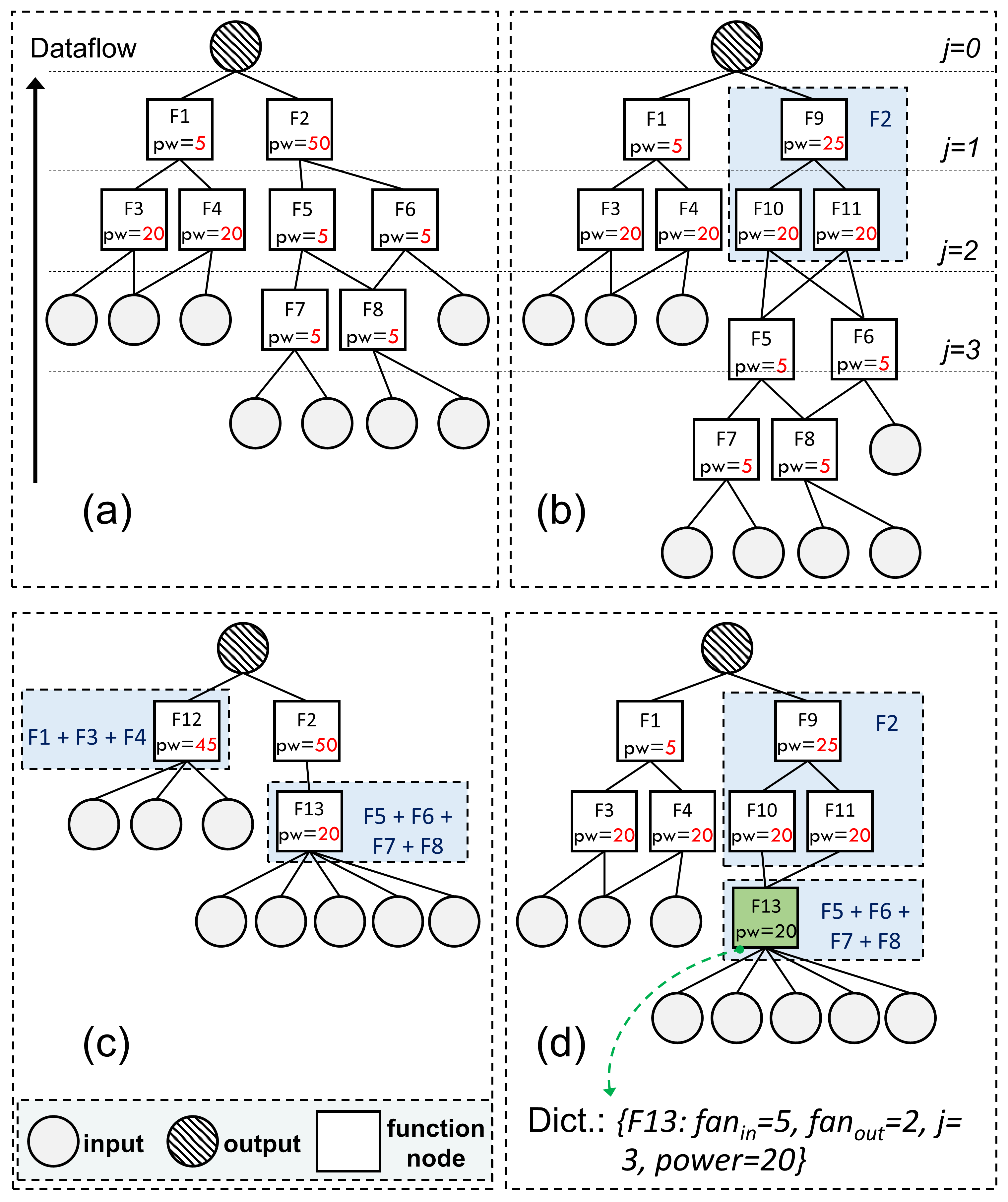}
\vspace{-0.75em}
    \caption{\small Tree illustrations of an 8-input/1-output design using different approaches (a) original, (b) Policy1, (c) Policy2, and (d) Policy3.}
    \vspace{-1.75em}
\label{tree}
\end{figure}

The Tree Generator takes the high-level program, in Step \encircle{1}, and synthesizes it to RTL-level Hardware Description Language (HDL), SPICE netlists, etc., and generates an un-optimized tree, where nodes contain functions and their power consumption, and edges indicate their connections. In Step \encircle{2}, we calculate power consumption using the commercial synthesis tool, including Synopsys DC and HSPICE.
Then, in Step \encircle{3}, the DIAC procedure will produce a \textsl{feature dictionary} (Dict.) and a \textsl{tree-based illustration}. Each node, e.g., node \emph{i} in level \emph{j} ($n^i_j$), has one feature dictionary, which contains the number of inputs from a lower level (fan\textsubscript{in}), the number of outputs to an upper level (fan\textsubscript{out}), the node level itself (\emph{j}), and its power consumption. 
A modified tree-based depiction of intermittency and peak harvested power ($V_{peak}$) will be derived using the following three strategies.
\emph{\ul{Policy1}}: Large components (functions) will be broken into smaller tasks with lower power to meet the desired conditions and criteria, including {\small $avg(F_{power})<V_{th}<<V_{peak}$}. Due to the division's nature, this provides the best resiliency at the cost of performance overhead. 
\emph{\ul{Policy2}}: Small components will be merged into larger components with a higher power to meet the conditions, {\small $max(F_{power})<<V_{th}$} and {\small $min(F_{power})=n\% Max$}, which provides best performance at the cost of lower resiliency, resulting from larger components. 
\emph{\ul{Policy3}}: This option is positioned between the first two policies and offers better performance and resiliency than Policies 1 and 2, respectively. 
Figure \ref{tree} depicts the original graph and its three representations regarding different conditions and needs.

\subsubsection{Replacement}
The replacement procedure aims to assess design feasibility and identify efficient replacement algorithms to address power failure, focusing on minimizing area and energy overheads.
Applying \textsl{Partitioning} and \textsl{NVM Insertion} modules to the produced tree ensures attestation function integrity and forward progress under various failures.
The partitioning is analogous to task-based methods existing in software approaches; however, our approach allows us to partition the netlist and ensure integrity statically. 
Given the modified tree from Step \encircle{4a}, power budget, and NVM features from Step \encircle{4b}, prioritizing nodes and finding replacement points efficiently requires weighing efficiency and resiliency. 
Generally, NVMs have a higher write cost than volatile memory, like SRAM, so reducing the number of NVMs' writes reduces power and delay. 
Three criteria define the replacement policy, realizing a more power-efficient system:
(I) \ul{Nodes in the upper level}: If NVMs are inserted in closer nodes to the output, the structure will be more power efficient under power failure conditions.
(II) \ul{Nodes with higher power consumption}: If NVMs are inserted in a node or a cone of nodes with a total higher power consumption, the system will be more power efficient.
(III) \ul{Nodes with higher $\text{fan}_{\text{in}}$ and/or $\text{fan}_{\text{out}}$}: If a node with more inputs and/or outputs is integrated with NVMs, the total number of writes will be reduced by a factor of $1/(\text{fan}_{\text{in}}$ + $\text{fan}_{\text{out}})$. 
Considering the above criteria, traversing the tree starts from leaves (bottom, inputs) upwards (roots, outputs) as follows:  
The total consumed power ({\small$P_{total}$}) until arriving at a node ($n$) equals the summation of all the previous nodes' power consumption.
By inserting an NVM in the position, $n$, the previous power values are set to zero, and the node's Dict. is updated with the new power consumption = {\small $P_{total}+P_n$}. Updates will be performed in parallel for all nodes at the same level. These criteria will be assessed and modified during the execution in Step \encircle{5}.

\subsubsection{Code generator}
After traversing all the levels and inserting NVMs, the NV-enhanced tree will be formed in \encircle{6}. 
Finally, in Step \encircle{7}, the generated code undergoes validation checks for possible timing violations.
To do so, this optimized tree will first be converted to an HDL code and submitted to the commercial tool. Upon passing, the design efficacy is determined based on the proposed performance metrics models and intermittency behavior.
Incorporating tree-based representations, different designs, and power failure scenarios will exponentially expand the design space. This will necessitate an efficient, precise, automated design tool that seamlessly converts any combinational and sequential designs into intermittent robust architectures without human intervention.
\vspace{-0.85em}

\subsection{Architectural States and Transitions}
\begin{figure}[t]
  \centering
    \includegraphics[width=1\linewidth]{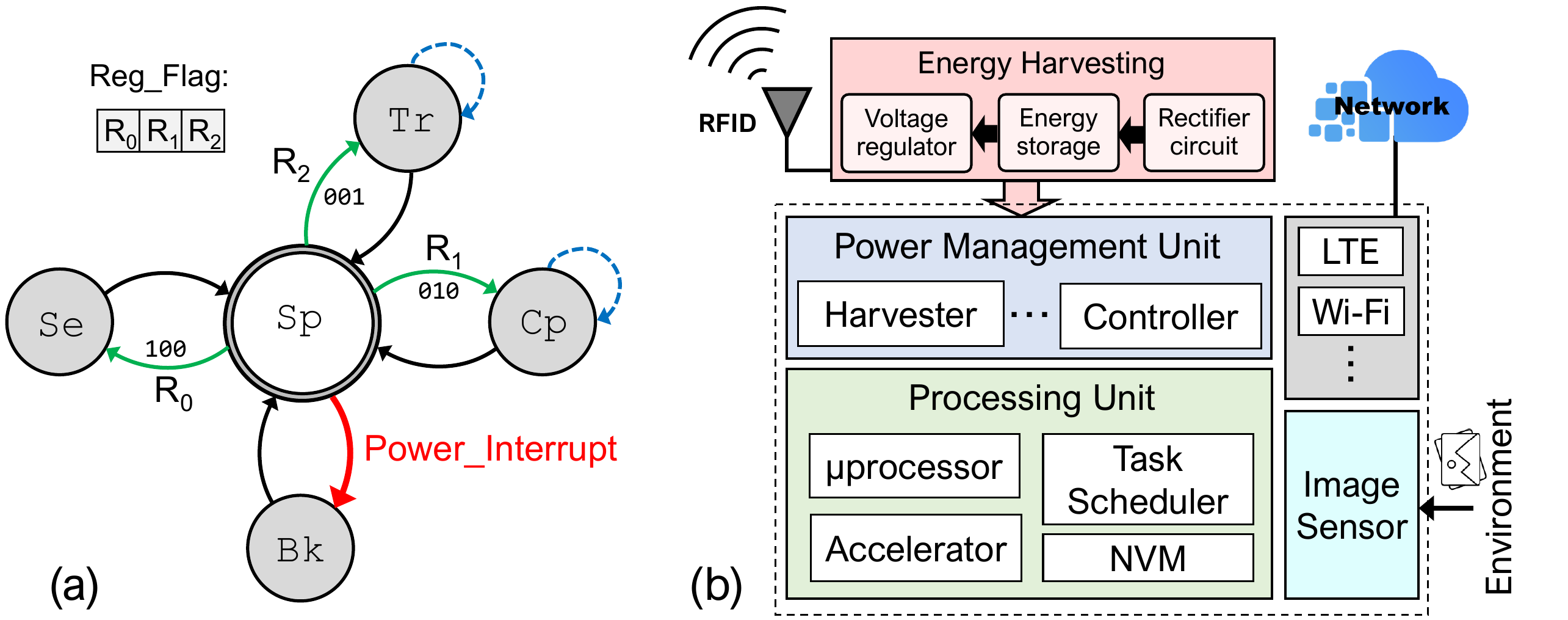}
\vspace{-1.75em}
    \caption{\small (a) The proposed state machine description, and (b) an intermittent-aware sensor node.}
 \vspace{0.5em}
\label{main}
\end{figure}

\vspace{-0.35em}
Batteryless systems can be broadly categorized into two primary types. The first category can complete all computations when the battery is fully charged. Conversely, the second type has inadequate capacitor energy storage to satisfy all computations.
For the second type, storing intermediate registers in NVMs is imperative. Herein, all operations, namely sense (\textsl{Se}), compute (\textsl{Cp}), transmit (\textsl{Tr}), sleep (\textsl{Sp}), and backup (\textsl{Bk}), are divided into atomic operations, which are executed uninterrupted.
These atomic operations are determined based on the system's maximum storage power and should only begin when sufficient power is available. We will iteratively use three policies to determine optimal atomic operations to maximize efficiency. The finite state machine (FSM) of the system offering the core operations of an IoT node, sense, compute, transmit, and sleep states is depicted in Fig. \ref{main} (a). Initially, the system is in \textsl{Sp} and reverts to this state after completing each atomic operation to conserve power for the next process. Integrating DIAC's produced design with the proposed FSM forms an intermittent-aware sensor node, ensuring resilience against power interruptions, shown in Fig. \ref{main}(b).

A detailed description of the required steps is illustrated in Algorithm \ref{Algorithm_gs}. 
The algorithm starts with the main while looping in line 3. This loop runs according to the \textsl{interval} value in line 33. As this loop is executed, the system's state changes depending on the Reg\_Flag and energy level (lines 6-11). 
The system has four threshold voltages for each state (\textsl{Th\textsubscript{State}}), e.g.,\textsl{Th\textsubscript{Cp}}, along with two more thresholds \textsl{Th\textsubscript{Safe\_Zone}} and \textsl{Th\textsubscript{Off}}. 
The system can perform tasks in each state if and only if criteria, including Reg\_Flag and enough power to complete the desired operation (\textsl{Energy$>$Th\textsubscript{State}}), are satisfied.
Once the system transitions to either the \texttt{\textbf{Cp}} or \texttt{\textbf{Tr}}, it will stay in this state until the power drops below the threshold defined by \textsl{Th\textsubscript{Safe\_Zone}} (lines 17 and 27). The \textsl{Safe\_Zone} threshold creates a narrow range that lies between the exit points of \texttt{\textbf{Cp}} or \texttt{\textbf{Tr}} and the beginning of \texttt{\textbf{Bk}} state.
The system can harvest further energy within this range, allowing it to restore to its prior state instead of reverting to \texttt{\textbf{Bk}}.
The transition to the next state is determined by modifying the value of the Reg\_Flag. However, this progression can be halted at any state by setting the Reg\_Flag to \texttt{`0b000'}.
For example, in \texttt{\textbf{Cp}}, the Reg\_Flag stays unchanged until the computation is done. Then, according to the computation result (lines 20-23), the Reg\_Flag is set to  \texttt{\textbf{Tr}} or  \texttt{\textbf{Sp}} and exits the loop.
During either \texttt{\textbf{Cp}} or \texttt{\textbf{Tr}}, when the power level falls below the \textsl{Safe\_Zone}, the system reverts to \texttt{\textbf{Sp}} (lines 25-32).
There are two interrupt routines in our system. The first interrupt is for the \texttt{Timer} (line 34), which is the maximum sampling rate of the system. However, it is important to note that this frequency can be reduced depending on the system's power. The second interrupt (\texttt{Power}) is generated by the power management unit (line 38). Suppose the total energy of the system is less than the threshold energy required for backup. In that case, the power management unit triggers an interrupt, and the backup unit stores all the necessary intermediate registers based on the register flag (Reg\_Flag). Since the energy of the system reduces in the \texttt{\textbf{Sp}} state (due to the standby energy), the system requires an interrupt to change its state to \texttt{\textbf{Bk}}. 
\vspace{-0.3em}


\setlength{\textfloatsep}{0pt}
\begin{algorithm}[!t] 
\scriptsize
\caption{\small State machine of the proposed system.}
\begin{algorithmic}[1]
\State \textbf{States}~=~[\texttt{\textbf{Sp}}, \texttt{\textbf{Se}}, \texttt{\textbf{Cp}}, \texttt{\textbf{Tr}}, \texttt{\textbf{Bk}}] \Comment{Sleep, Sense, Compute, Transmission, and Backup.}
\State \texttt{\textbf{Initialize}} State $\leftarrow$ \texttt{\textbf{Sp}}
\State\texttt{\textbf{while}}~(True)
\State\hspace{0.25cm}\texttt{\textbf{if}} (State == \texttt{\textbf{Sp}})
\State\hspace{0.25cm}\hspace{0.25cm}Reg $\leftarrow$ \texttt{\textbf{Read}} (Reg\_Flag)
\State\hspace{0.25cm}\hspace{0.25cm}\texttt{\textbf{if}} (Reg == \texttt{0b100} $\&$ energy $>$ Th\textsubscript{Se})
\Comment{The \textcolor{mygreen}{green} arrow from Sp to Se.}
\State\hspace{0.25cm}\hspace{0.25cm}\hspace{0.25cm}State $\leftarrow$ \texttt{\textbf{Se}}
\State\hspace{0.25cm}\hspace{0.25cm}\texttt{\textbf{elsif}} (Reg == \texttt{0b010} $\&$ energy $>$ Th\textsubscript{Cp}) 
\Comment{The \textcolor{mygreen}{green} arrow from Sp to Cp.}
\State\hspace{0.25cm}\hspace{0.25cm}\hspace{0.25cm}State $\leftarrow$ \texttt{\textbf{Cp}}
\State\hspace{0.25cm}\hspace{0.25cm}\texttt{\textbf{elsif}} (Reg == \texttt{0b001} $\&$ energy $>$ Th\textsubscript{Tr}) 
\Comment{The \textcolor{mygreen}{green} arrow from Sp to Tr.}
\State\hspace{0.25cm}\hspace{0.25cm}\hspace{0.25cm}State  $\leftarrow$ \texttt{\textbf{Tr}}
\State\hspace{0.25cm}\texttt{\textbf{if}} (State == \texttt{\textbf{Se}})
\State\hspace{0.25cm}\hspace{0.25cm}Sample $\leftarrow$ \texttt{\textbf{Sense()}}
\State\hspace{0.25cm}\hspace{0.25cm}State $\leftarrow$ \texttt{\textbf{Sp}}
\State\hspace{0.25cm}\hspace{0.25cm}Reg $\leftarrow$ \texttt{0b010}
\State\hspace{0.25cm}\texttt{\textbf{if}} (State == \texttt{\textbf{Cp}}) 
\State\hspace{0.25cm}\hspace{0.25cm}\texttt{\textbf{while}}( energy $>$ Safe\_Zone) \Comment{Dashed \textcolor{blue}{blue} arrow in Cp.}
\State\hspace{0.25cm}\hspace{0.25cm}\hspace{0.25cm}Result $\leftarrow$ \texttt{\textbf{Compute(}}Sample\texttt{\textbf{)}}
\State\hspace{0.25cm}\hspace{0.25cm}\hspace{0.25cm}\texttt{\textbf{if}} (sample is completely processed) 
\State\hspace{0.25cm}\hspace{0.25cm}\hspace{0.25cm}\hspace{0.25cm}\texttt{\textbf{if}} (transmission is require)
\State\hspace{0.25cm}\hspace{0.25cm}\hspace{0.25cm}\hspace{0.25cm}\hspace{0.25cm}Reg $\leftarrow$ \texttt{0b001}
\State\hspace{0.25cm}\hspace{0.25cm}\hspace{0.25cm}\hspace{0.25cm}\texttt{\textbf{else}}
\State\hspace{0.25cm}\hspace{0.25cm}\hspace{0.25cm}\hspace{0.25cm}\hspace{0.25cm}Reg $\leftarrow$ \texttt{0b000}
\State\hspace{0.25cm}\hspace{0.25cm}\hspace{0.25cm}\hspace{0.25cm}\texttt{\textbf{Break}} \Comment{Break the loop!}
\State\hspace{0.25cm}\hspace{0.25cm}State $\leftarrow$ \texttt{\textbf{Sp}} \Comment{The \textcolor{black}{black} arrow from Cp to Sp.}
\State\hspace{0.25cm}\texttt{\textbf{if}} (State == \texttt{\textbf{Tr}}) 
\State\hspace{0.25cm}\hspace{0.25cm}\texttt{\textbf{while}}(energy $>$ Safe\_Zone)\Comment{Dashed \textcolor{blue}{blue} arrow in Tr.}
\State\hspace{0.25cm}\hspace{0.25cm}\hspace{0.25cm}\texttt{\textbf{Transmit}} (Result)
\State\hspace{0.25cm}\hspace{0.25cm}\hspace{0.25cm}\texttt{\textbf{if}} (transmission is done) 
\State\hspace{0.25cm}\hspace{0.25cm}\hspace{0.25cm}\hspace{0.25cm}Reg $\leftarrow$ \texttt{0b000}
\State\hspace{0.25cm}\hspace{0.25cm}\hspace{0.25cm}\hspace{0.25cm}\texttt{\textbf{Break}}\Comment{Break the loop!}
\State\hspace{0.25cm}\hspace{0.25cm}State $\leftarrow$ \texttt{\textbf{Sp}} \Comment{The \textcolor{black}{black} arrow from Tr to Sp.}
\State\hspace{0.25cm}\texttt{\textbf{Sleep}} (interval)
\Comment{Interval is determined by the average charging rate.}
\State \texttt{\textbf{interrupt\_Timer}}() \Comment{Check the sensing \textcolor{red}{interval} is valid or not.}
\State \hspace{0.25cm}Reg $\leftarrow$ Read\_Reg\_Flag (interval)
\State \hspace{0.25cm}\texttt{\textbf{if}} (Reg == \texttt{0b000})
\State \hspace{0.25cm}\hspace{0.25cm}Reg $\leftarrow$ \texttt{0b100}
\State \texttt{\textbf{interrupt\_power}}() \Comment{\textcolor{red}{Power is scarce} to perform any task!}
\State \hspace{0.25cm}State $\leftarrow$ \texttt{\textbf{Bk}}
\State \hspace{0.25cm}\texttt{\textbf{Backup()}}\Comment{Back up all intermediate values w.r.t. register value.}
\State \hspace{0.25cm}State $\leftarrow$ \texttt{\textbf{Sp}}\Comment{The \textcolor{black}{black} arrow from Bk to Sp.}

\end{algorithmic}
\label{Algorithm_gs}
\end{algorithm}

\vspace{-0.3em}
\section{Results}
\vspace{-0.5em}
\subsection{Validation of DIAC Approach}
As previously noted, \textsl{Policy3} simultaneously provides acceptable resiliency and efficiency. Thus, we utilize it as a replacement policy and the principal methodology in this section. Figure \ref{tree}(d) shows an example of this approach where the upper and lower limits for power consumption are set at $25$mJ and $20$mJ per operand, respectively.
Consequently, operands consuming more than $25$mJ of power must be divided, while those using less than $20$mJ should be combined. Thus, F$_{5-8}$ are merged to be represented by F$_{13}$. Conversely, F$_2$ is broken down into smaller operands, i.e., F$_{9-11}$. This step is followed by creating a dictionary based on the operands' number of inputs and outputs, delays, and power consumption. 
To accurately estimate the power and delay of each operation, DIAC harnesses the output from HSPICE, considering delay, static, and dynamic power for each operand, which comprises multiple gates.
We then devised a mathematical model to meet diverse needs and estimate design-time evaluation parameters before run-time.
We approximated the dynamic energy using {\small $\approx 2 \times  \sum_{i=0}^{n}delay_i \times dynamic\_power_i$}, where $n$ represents the total number of gates in an operand, and the delay is determined when both input and output equal ${V_{DD}}\over{2}$. For a more accurate energy consumption estimation, this delay is doubled. It is essential to highlight that while one gate is in switching mode, other gates remain inactive without any activities. Therefore, the power consumption of these inactive gates is determined by the critical delay path (CDP) multiplied by the total number of gates ({\small $\approx CDP \times \sum_{i=0}^{n-1} static\_power_i$}), excluding the currently active gate.
A system-level in-house framework is developed to verify our approach's functionality. 
Firstly, we implement a design by traversing its NV-enhanced tree generated by DIAC and concerning the system's parameters, including peripherals, using HSPICE in the 45nm NCSU PDK library.
The memory controller and registers are designed and synthesized by Design Compiler. Afterward, we incorporated the results from circuit-level assessments and extensively modified CACTI at the architecture level. Next, we integrated the architecture with the proposed FSM and exported the performance to an in-house cross-layer framework, taking the CACTI output and application netlist as the inputs. At the application, we evaluated the performance of our proposed technique in the presence of power outages using various benchmarks. 
To incorporate the behavior of intermittency into our study, we adopted a methodological approach that involved simulating an intermittent power source characterized by a predetermined sequence of voltage levels that cyclically repeat. To accurately represent the behavior of intermittent power, we introduced a virtual energy source within our simulation framework, designed to mimic the functionality of a battery. This virtual energy source is responsible for accumulating energy during power availability and deducting energy consumption during periods of power unavailability.
Throughout the simulation process, we closely monitored several key parameters to ensure an accurate representation, including power/voltage levels, power availability, and the level of the virtual energy source. Monitoring is achieved by observing the behavior of the power source and virtual battery, as well as tracking the power consumption associated with various computational operations. 
In the system, a capacitance of $2$mF is considered, and an operational voltage of 5V is used. Therefore, the system can store a maximum of E\textsubscript{MAX} $=25$mJ of energy.
Herein, we assume that the sense, compute, and transmit operations consume $2$mJ, $4$mJ, and $9$mJ, respectively, all with a $\pm \%10$ uncertainty.
Furthermore, the \textsl{Th\textsubscript{Safe\_Zone}} region exceeds the backup threshold by $2$mJ. In this zone, the system enters sleep mode without requiring a backup. However, if the battery energy drops below \textsl{Th\textsubscript{Bk}}, the NVMs save all essential registers.
\begin{figure}[t]
  \centering
    \includegraphics[width=\linewidth]{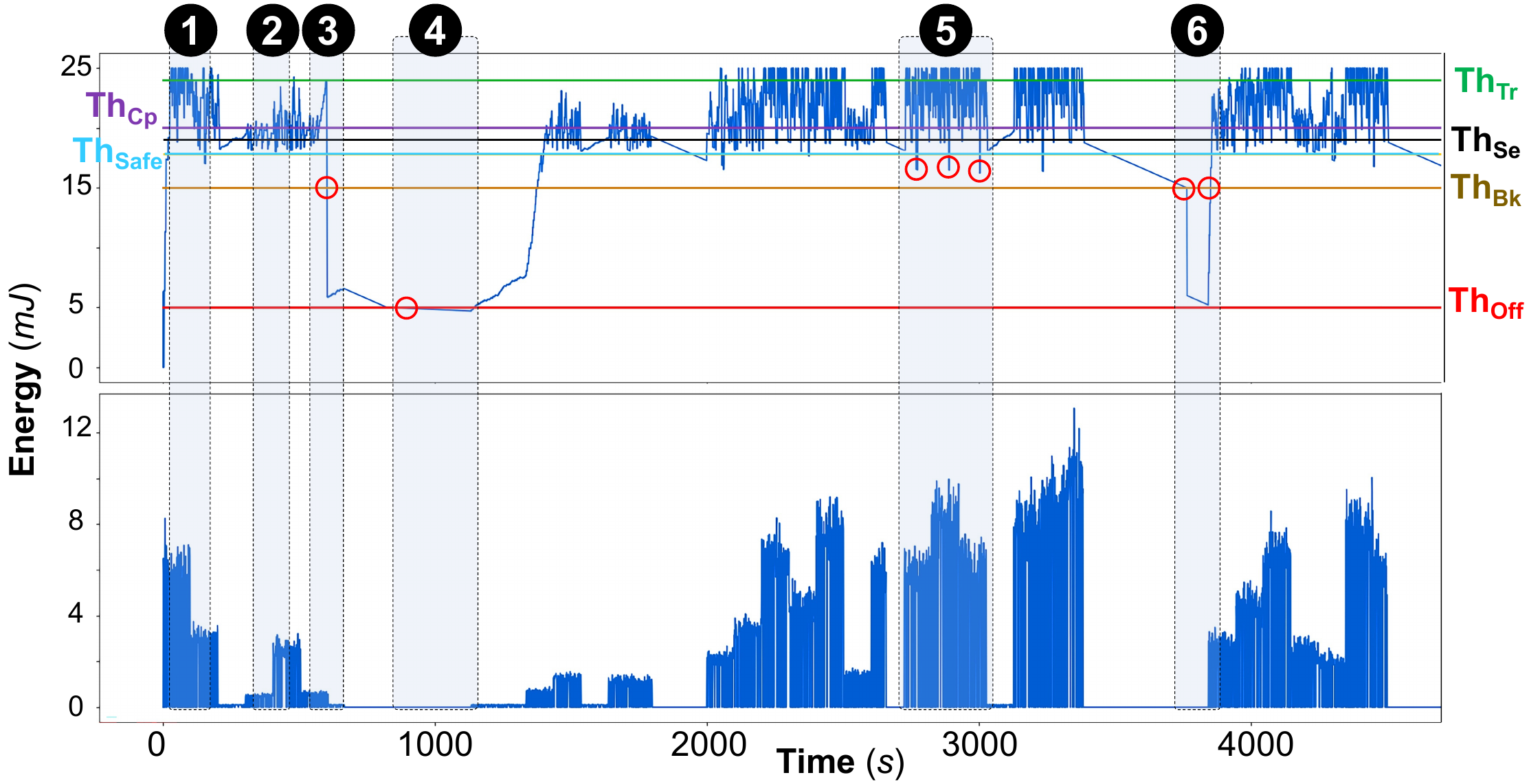}
\vspace{-2em}
    \caption{\small Battery (E\textsubscript{Batt}) (top) and charging rate (bottom) of the system.}
\vspace{0.5em}
\label{charge}
\end{figure}

Figure \ref{charge} illustrates the energy (E\textsubscript{Batt}) stored in the capacitor concerning the system's charging rate.
In \encircle{1}, the charging rate surpasses the system's needs. Consequently, the energy stored occasionally reaches its maximum capacity of E\textsubscript{MAX}, i.e., $25$mJ. This allows the system to operate at peak performance.
Conversely, in \encircle{2}, the charging rate is insufficient to meet the system's demands. The system remains in the \texttt{\textbf{Sp}} state until E\textsubscript{Batt} surpasses \textsl{Th\textsubscript{Cp}}. It then transitions to either \texttt{\textbf{Cp}} or \texttt{\textbf{Tr}} states. During this phase, the system continues operations until the energy drops below \textsl{Th\textsubscript{Safe\_Zone}}.
In \encircle{3}, a sudden decline in the charging rate is noted, falling below \textsl{Th\textsubscript{Bk}}. Consequently, the system backs up its registers to the NVMs. 
Following the backup in \encircle{4}, a sustained low charging rate causes E\textsubscript{Batt} to drop beneath \textsl{Th\textsubscript{Off}}, resulting in a complete system shutdown. Upon accumulating sufficient power, the system then retrieves register data based on the NVM values.
As previously explained, the \textsl{Th\textsubscript{Safe\_Zone}} threshold is crucial in minimizing NVM writes. As exemplified in \encircle{5}, the system enters this zone thrice, maintaining its state in \texttt{\textbf{Sp}}. Throughout these three instances, since E\textsubscript{Batt} never fell below \textsl{Th\textsubscript{Bk}}, no energy-intensive NVM writes were needed. Subsequent efficient energy harvesting allowed the system to transition back to an active domain, fetching states directly from volatile storage and the Reg\_Flag.
A different scenario unfolds in \encircle{6}, where the charging source is interrupted, prompting the system to revert to the \texttt{\textbf{Sp}} state. Despite being in this state, a minimal leakage current persists, causing the system's energy to fall below \textsl{Th\textsubscript{Bk}}. This triggers a backup process. But the charging is restored before a complete power outage, enabling the system to resume operations. Herein, there's no necessity to fetch register values from the NVMs.
\begin{figure*}[t]
  \centering
    \includegraphics[width=\linewidth]{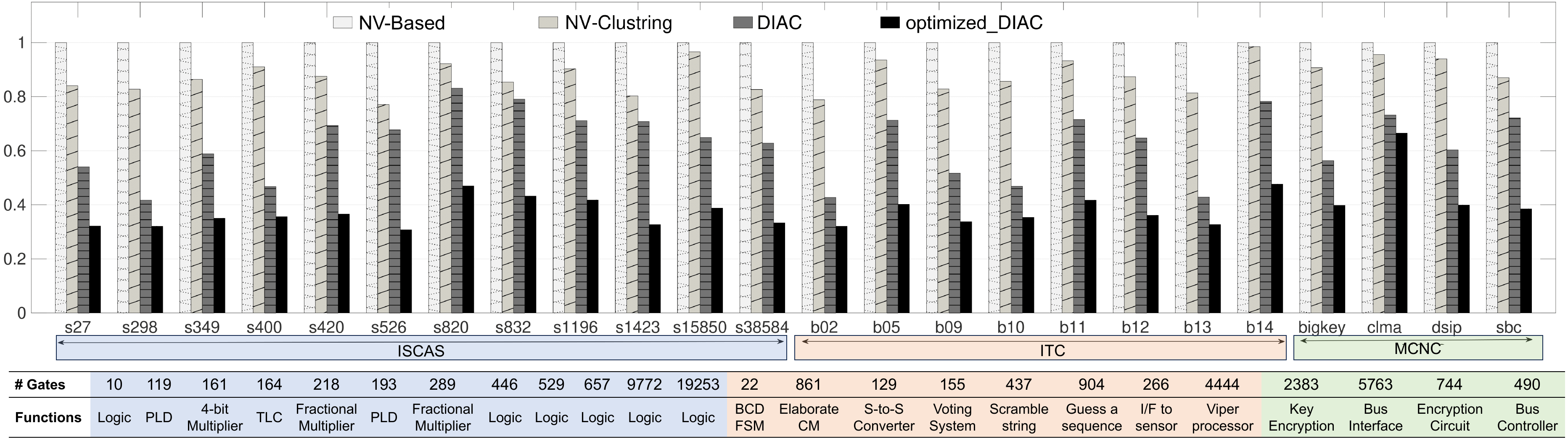}
\vspace{-1.75em}
    \caption{\small Normalized PDP compared to intermittent-aware implementations for different benchmark circuits.}
    \vspace{-1em}
\label{pdp}
\end{figure*}

\subsection{Performance Evaluation}
In this section, the developed design tool, including DIAC and the FSM, is leveraged to implement large-scale circuit benchmarks, including ISCAS-89, ITC-99, and MCNC.
To evaluate DIAC performance using the developed framework, the power-delay-product (PDP) values for the four schemes are considered. The NV-based method operates similarly to conventional checkpointing, where flip-flops (FFs) are replaced by the NV-FFs to store states. It provides the highest resiliency at the cost of significant overhead. The second compared technique is NV-Clusting, presented in \cite{Roohi2018TC}. In this approach, the authors introduced a new concept, logic-embedded flip-flop (LE-FF), which can realize Boolean logic functions and an inherent state-holding capability. 
To make the evaluation more comprehensive, we have considered two DIAC-based implementations, excluding and including \textsl{Th\textsubscript{Safe\_Zone}}, denoting DIAC, and optimized\_DIAC designs, respectively.
According to subsection 2.B, this state allows us to reduce power consumption and delay by reducing the number of NVM writes required.
Because of this, the optimized\_DIAC process significantly reduces the number of NVM writes as a costly operation. It is worth noting that the safe zone varies based on the harvested energy.
Figure \ref{pdp} depicts the PDP results for all the mentioned approaches.
For various ISCAS-89, ITC-99, and MCNC benchmark circuits, these results exhibit an average of 36\% (25\%), 41\% (33\%), and 34\% (28\%) PDP improvements, respectively, for the DIAC\_based designs compared to NV-based (NV-clustering) implementations. Further PDP improvements are achieved by using the optimized\_DIAC methodology, which provides up to 61, 56, and 38 percent average PDP improvements compared to NV-based, NV-clustering, and DIAC approaches, respectively. The benefits are achieved because of the optimal NVM write operations.

\subsection{Conditions and Discussion}
Two assumptions are considered when assessing the DIAC methodology:(1) There is never enough energy in the system to complete a process (instance). This means that using conventional CMOS-based designs without NVMs will not work. To ensure this condition, if a benchmark circuit, e.g., s27, consumes less energy than battery capacity, it is rerun multiple times till the total energy exceeds the capacity. Then, we considered all performed operations as one bigger and/or more complex task; 
(2) To make a fair comparison among different intermittent robust computing systems, the same NVM technology is leveraged. 
Extensive research has been conducted on designing NVMs using different NV elements, such as magnetic RAM (MRAM), resistive random access memory (ReRAM), ferroelectric random-access memory (FeRAM), and phase change memory (PCM). 
Due to the International Technology Roadmap for Semiconductors (ITRS) report, which identifies
spintronic devices as capable post-CMOS candidates, we chose MRAM as our NVM technology. MRAM cells provide
non-volatility, near-zero standby power, high integration density, and radiation-hardness features. Moreover,
MTJs can be vertically integrated at the back-end CMOS fabrication process, resulting in lower interconnect
energy losses and less area overhead. 
It is noteworthy that although varying NVM technology changes (reduces/increases) the enhancement, the overall improvement trend remains relatively stable. This is achieved because the DIAC approach optimizes NVM writes as an energy-hungry operation. For example, if ReRAMs replace MRAM cells, the optimized\_DIAC exhibits higher efficiency than the other examined techniques because the ReRAM write consumes $\sim 4.4\times$ more energy than MRAM.








\section{Conclusion}
This paper proposed the \texttt{\textbf{DIAC}} methodology to enhance intermittent computing systems' efficiency. This synthesis process is intricately designed to ensure the progress of the tasks while optimizing energy consumption. Coupled with a detailed FSM that characterizes core IoT operations, the DIAC-based design proved efficiency and resiliency against power disruptions. Furthermore, a wide range of benchmark circuits has showcased the DIAC's superiority over the previous schemes, with a significant PDP improvement. 





\section*{Acknowledgments}
This work is supported in part by the National Science Foundation under Grant No. 2303114.

\bibliographystyle{plain}
\bibliography{main}
\end{document}